\documentclass[reprint,amsmath,amssymb,aps,prl,floatfix,raggedbottom,footinbib]{revtex4-2}
\usepackage{array}
\usepackage{subfigure}
\usepackage{graphicx}
\usepackage{dcolumn}
\usepackage{bm}
\usepackage{xcolor}
\usepackage[colorlinks=true,allcolors=blue]{hyperref}

\newcommand{\bk}{\mathbf{k}}
\newcommand{\bq}{\mathbf{q}}

\newcommand{\br}{\mathbf{r}}
\newcommand{\mcA}{\mathcal A}

\begin{document}

\title{Quantum-Metric-Driven Hollow Wigner Crystals}
\author{Zekun Zhuang}
\affiliation{Institute for Advanced Study, Shenzhen University, Shenzhen 518060, China}

\begin{abstract}
Recent studies have revealed the profound influence of Berry curvature on electron crystallization, yet the independent role of the quantum metric remains unclear. Here, we introduce a minimal three-band continuum model with separately tunable quantum metric and Berry flux. 
Employing the Hartree-Fock approach, we find that a conventional s-wave Wigner crystal can develop a hollow charge distribution with depleted density at lattice sites, without a change in symmetry or topology. We show that this density reconstruction occurs even at vanishing Berry curvature and arises from strong quantum metric and the nontrivial angular structure of the form factor. In our model, the continuous crossover between hollow and conventional WCs is accompanied by anomalous direct exchange interactions, nearly vanishing at intermediate quantum metric but enhanced by over an order of magnitude in the strong-metric regime. These results reveal how quantum metric can reshape both the charge distribution and exchange interactions of electron crystals beyond simple wave-function delocalization.
\end{abstract}

\maketitle
\textit{Introduction.---}
Quantum geometry (QG), encoded in the momentum dependence of Bloch wave functions, has emerged as a fundamental ingredient in understanding quantum materials beyond their energy dispersions \cite{Yu2025,Gao2026,Liu2025}. Its two components, Berry curvature (BC) and quantum metric (QM), characterize the geometric phase and quantum distance between electronic states, respectively. While BC underlies anomalous transport and topological phenomena \cite{Di2010,Bansil2016,Qi2011}, QM plays an essential role in a variety of physical effects, including the superfluid stiffness and coherence length of flat-band superconductors \cite{Peotta2015,Chen2024,Hu2025} and nonlinear electronic and optical responses \cite{Kaplan2024,Avdoshkin2025}.

In particular, recent studies have revealed a profound influence of BC on electron crystallization, motivated in part by experimental signatures of anomalous Hall crystals (AHCs) in multilayer graphene and moir\'e heterostructures \cite{Lu2025,Su2025,Waters2025,Liu2025_2,Li2026}. Being odd under time reversal, BC introduces a natural source of chirality into electronic states. Unlike conventional Wigner crystals (WCs), anomalous Hall crystals (AHCs) carry a nonzero Chern number while spontaneously breaking continuous translational symmetry \cite{Tan2024,Tan2025,Dong2024,Soejima2024,Zeng2024,Soejima2025}. Moreover, sufficiently large Berry flux can favor unconventional WCs whose localized electrons acquire nonzero internal orbital angular momentum \cite{Soejima2025,Joy2025,Desrochers2026,Zverevich2026}. These findings demonstrate how BC can endow electron crystals with nontrivial topology and orbital chirality, giving rise to crystalline states beyond the conventional Wigner picture.

By contrast, how QM independently affects electron crystallization in the absence of BC remains largely unexplored. In the conventional harmonic picture of WCs, electrons form centrally peaked Gaussian-like wave packets, and QM, known to impose a lower bound on their spatial spread \cite{Marzari1997,Souza2000}, might naively be expected to simply broaden these wave packets. Whether QM can instead qualitatively reshape their density profiles remains unclear.  Previous theoretical studies have often adopted continuum toy models with ideal QG, ranging from parent bands with uniform BC to models with nonuniform curvature profiles but quantized total Berry flux \cite{Tan2024,Soejima2025,Desrochers2026}. In these models, the QM is tied to BC, making their respective effects difficult to disentangle. Moreover, BC in realistic materials is generally momentum-dependent, and the Berry flux within a low-energy region need not be quantized, unlike the total flux over the Brillouin zone. These limitations motivate a minimal continuum model with separately tunable QM strength and Berry flux, enabling the study of electron crystallization under more general conditions while disentangling their distinct effects.

In this Letter, we construct a minimal three-band continuum model with separately tunable Berry flux and quantum-metric strength. Within self-consistent Hartree-Fock (HF) theory, we map out the crystalline phase diagram, focusing on the previously unexplored regime of vanishing Berry curvature. Remarkably, we find that increasing QG strength drives a centrally-peaked s-wave WC to a hollow WC (HWC) with centrally-depleted density, even in the complete absence of BC. We derive an analytical criterion for hollowing from the angular structure of the band form factor, revealing its microscopic origin. We also uncover strikingly nonmonotonic direct Coulomb exchange, nearly suppressed near the crossover by destructive inter-channel interference but enhanced by over an order of magnitude in the strong-metric regime. Together, these findings reveal a previously overlooked role of quantum geometry in reorganizing the internal structure and exchange interactions of otherwise conventional electron crystals.

\begin{figure*}[t]
 \centering
 \includegraphics[width=0.32\textwidth]{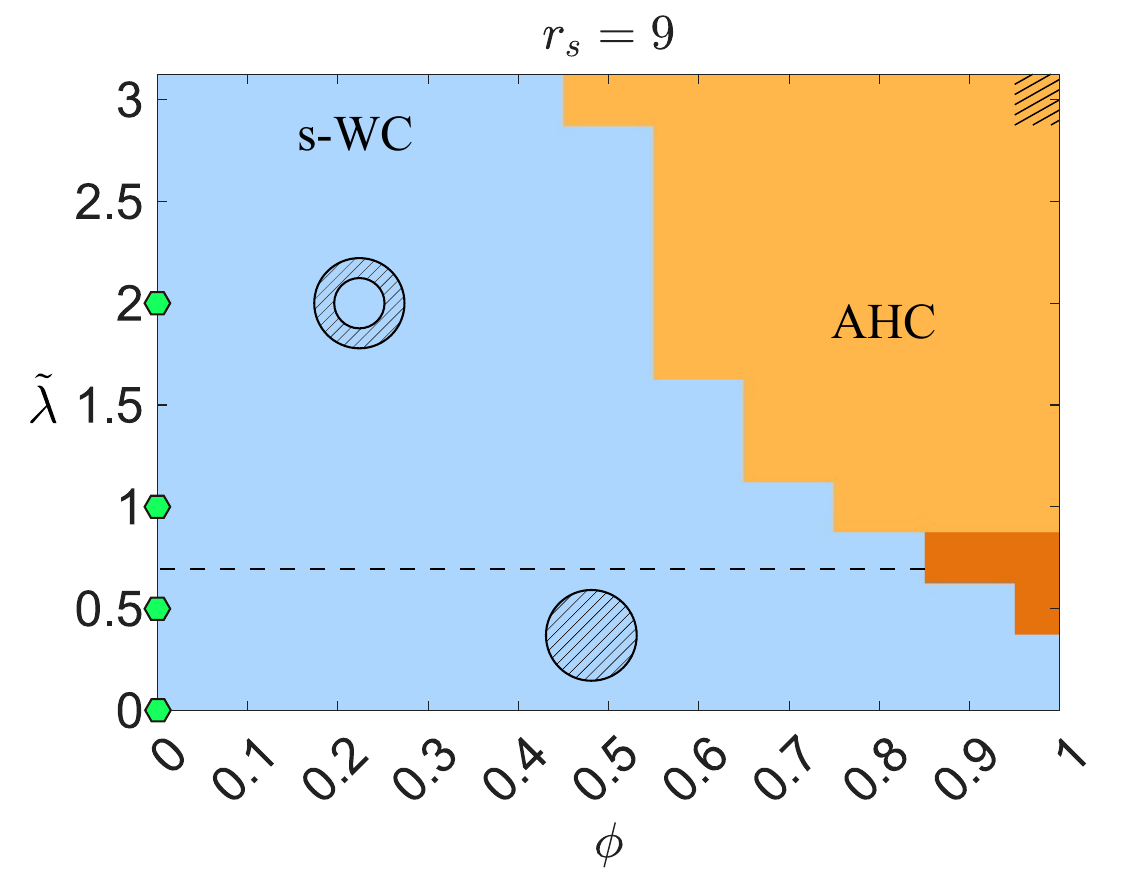}
 \includegraphics[width=0.32\textwidth]{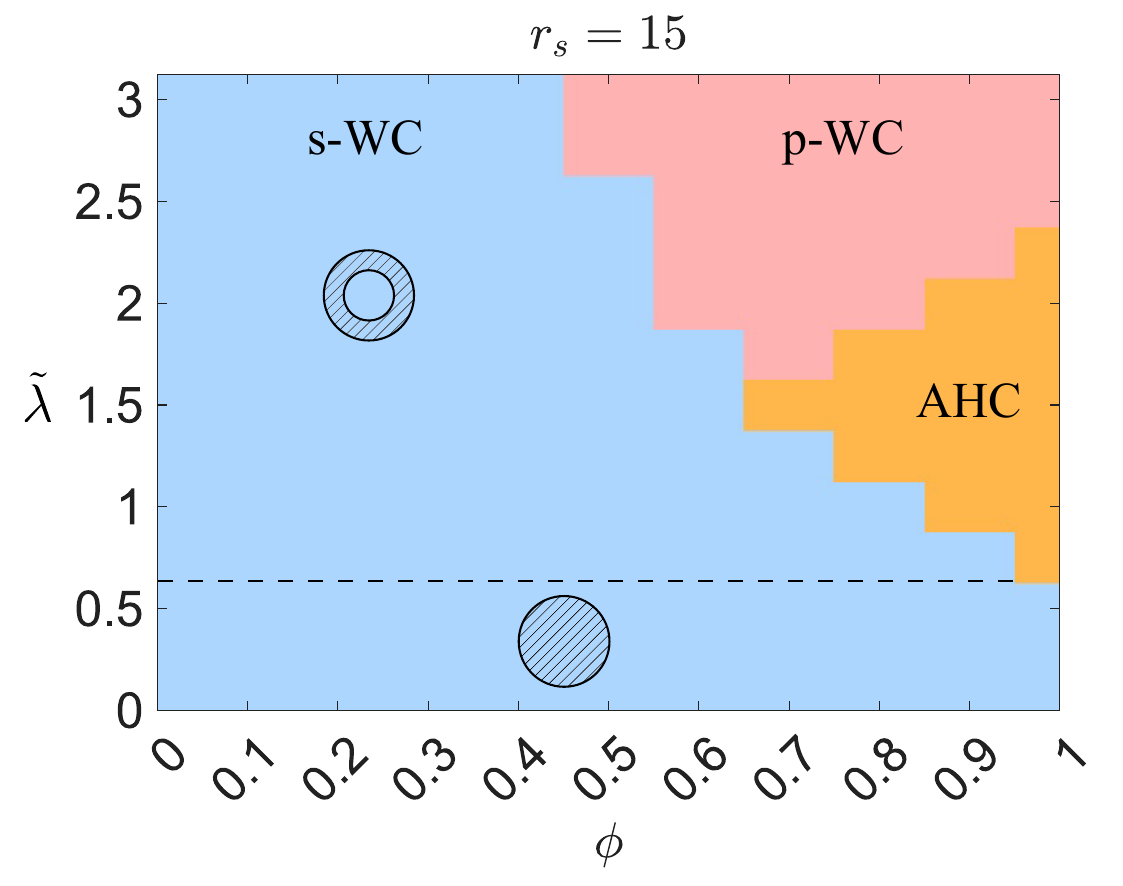}
 \includegraphics[width=0.32\textwidth]{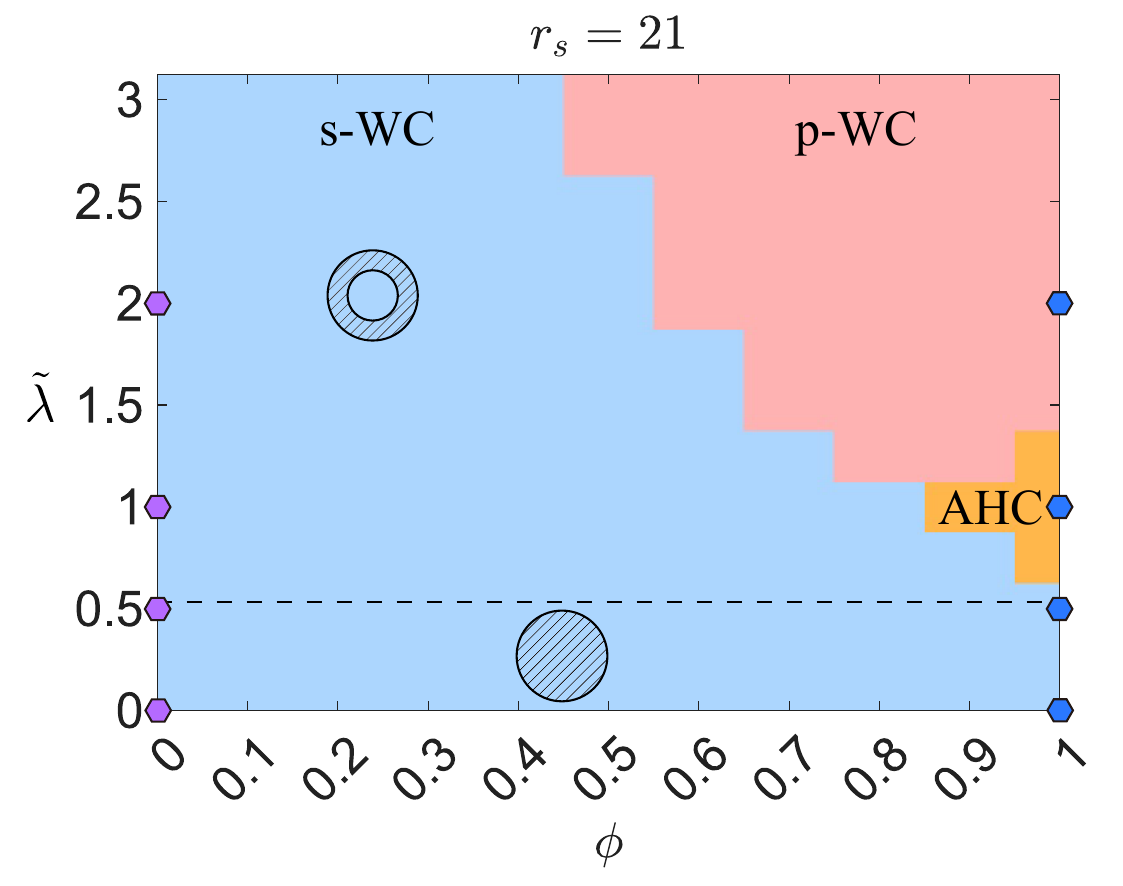}
 \caption{The representative phase diagram in the $\tilde\lambda-\phi$ plane at (a) $r_s=9$ (b) $r_s=15$ (c) $r_s=21$. The blue and pink regions represent the triangular s-WC and p-WC, while the yellow and orange regions represent the triangular and square AHC, respectively. The dashed line is a schematic crossover between hollow s-WC (concentric circles) and regular s-WC (single circle). The shaded region represents a possible metallic phase \cite{MetallicPhaseNote}.}
 \label{fig:phasediagram}
\end{figure*}

\textit{Three-band continuum model.---}
We consider the following three-component continuum Hamiltonian in two dimensions ($\hbar=1$)
\begin{equation}
 h_0(\bk)=\frac{k^2}{2m}\mathbb{I}+
 \Delta\begin{pmatrix}
 \lambda^2k^2&a k_-&b k_+\\
 a k_+&1&0\\
 b k_-&0&1
 \end{pmatrix},
 \label{eq:h0}
\end{equation}
where $k_\pm=k_x\pm ik_y$ and $\Delta>0$. The lowest band has standard parabolic dispersion
$\epsilon_{\bk}^0=k^2/(2m)$ and the corresponding normalized eigenvector is
\begin{equation}
 |u_{\bk}^0\rangle=
 \frac{(1,-a k_+,-b k_-)^T}{\sqrt{1+\lambda^2k^2}}.
 \label{eq:spinor}
\end{equation}
The other two high-energy bands lie at $\epsilon_{\bk}^1=\epsilon_{\bk}^0+\Delta$ and
$\epsilon_{\bk}^2=\epsilon_{\bk}^0+\Delta(1+\lambda^2k^2)$, and can be projected out in the infinite gap limit $\Delta\rightarrow\infty$. It proves convenient to parametrize $a=\lambda\sqrt{(1-\phi)/2}$ and $b=\lambda\sqrt{(1+\phi)/2}$, with $\lambda\geq0$ and
$-1\leq\phi\leq1$.  The QG of the lowest band $|u_{\bk}^0\rangle$ is then controlled by $\lambda$ and $\phi$, which determine its characteristic momentum scale and total Berry flux, respectively. With $\bm{\mcA}=i\langle u_{\bk}|\bm\nabla_{\bk}u_{\bk}\rangle$, the BC is
\begin{equation}
 \Omega(k)=\frac{2\phi\lambda^2}{(1+\lambda^2k^2)^2},
 \qquad \int d^2k\,\Omega(k)=2\pi\phi.
 \label{eq:berry}
\end{equation}
The radial and tangential metric components are
\begin{align}
 g_\parallel(k)&=\frac{\lambda^2}{(1+\lambda^2k^2)^2},\nonumber\\
 g_\perp(k)&=\frac{\lambda^2[1+(1-\phi^2)\lambda^2k^2]}
 {(1+\lambda^2k^2)^2}.
 \label{eq:metric}
\end{align}
When $\phi=0$, the BC vanishes everywhere while the metric remains finite; when $|\phi|=1$ this three-band $\lambda-\phi$ model becomes block diagonal and reduces to the two-band $\lambda$-Jellium model \cite{Soejima2025}. We emphasize that this continuously tunable total flux is allowed because the continuum model lives on a noncompact infinite momentum plane, thus it can only appear as a low-energy effective theory in real solid systems.

Assuming the gap $\Delta$ is infinite, projecting a Coulomb interaction onto the lowest band $|u_{\bk}^0\rangle$ gives the low-energy interacting (spinless) Hamiltonian
\begin{equation}
 H=\sum_{\bk}\epsilon_{\bk}c^\dagger_{\bk}c_{\bk}
 +\frac{1}{2A}\sum_{\bq\ne0}V(q):\widetilde\rho_{\bq}
 \widetilde\rho_{-\bq}:.
 \label{eq:projectedH}
\end{equation}
Here the momentum-space Coulomb interaction is $V(q)=2\pi/|q|$ (we have set $\epsilon=e=1$) and $A$ is the system area. The projected density operator is
\begin{equation}
   \widetilde\rho_{\bq}=\sum_{\bk}F(\bk,\bk+\bq)
 c^\dagger_{\bk}c_{\bk+\bq},\label{eq:projecteddensity}  
\end{equation}
in which the form factor is given by
\begin{equation}
  F(\bk,\bk')=\langle u_{\bk}^0| u_{\bk'}^0\rangle =
 \frac{1+\lambda^2\bk\!\cdot\!\bk'
 -i\phi\lambda^2(\bk\!\times\!\bk')_z}
 {\sqrt{(1+\lambda^2k^2)(1+\lambda^2k'^2)}}.
 \label{eq:overlap}   
\end{equation}
We note that a Jellium background has been implicitly added to remove the electrostatic energy divergence.

After projection, the system is parametrized by three independent dimensionless parameters $r_s=a_0/a_B$, $\tilde\lambda=\lambda/a_0$ and $\phi$, in which $a_0=(\pi n)^{-1/2}$ is the interparticle distance and $a_B=m^{-1}$ is the Bohr radius. The energy is represented in units of Hartree, which equals $m$ in our convention. In the following, we mainly consider $\phi \geq 0$, while the opposite case can be easily inferred. As the Bloch wavefunction $|u_{\bk}^0\rangle$ varies smoothly in momentum space when $\tilde \lambda\ll1$ but drastically when $\tilde \lambda\gg1$, we denote the former as the \emph{weak QG} regime while the latter as the \emph{strong QG} regime.

\textit{Crystalline phase diagram.---}
The HF phase diagrams in the $\lambda-\phi$ plane at representative $r_s$ are shown in Fig. \ref{fig:phasediagram}(see End Matter for the method). We find that at selected parameters, crystals always have lower energy than the liquid. Different crystalline phases can be classified according to their rotation eigenvalue at high-symmetry points \cite{Fang2012}(see End Matter for more details), provided the discrete rotation symmetry is preserved. There are mainly three distinct phases: s-wave WC (s-WC) and p-wave WC (p-WC) which are topologically trivial, and the AHC which has nonzero Chern number ($|C|=1$ here). The s-WC is the standard WC which acquires no phase upon consecutive lattice rotation, revealing its s-wave character; in comparison,  the p-WC, also known as the `Halo WC' or `chiral WC' in recent works \cite{Soejima2025,Joy2025,Desrochers2026},  acquires a $2\pi/n$ phase per electron under $C_n$ rotation and thus has a $p$-wave character in its momentum-space wavefunction.

Our results at $|\phi|=1$ recover most of the features obtained in Ref. \cite{Soejima2025}, including the topological transition between s-WC and AHC, attributed to the closing and reopening of a topological single-particle gap. Away from $|\phi|=1$, the region of AHC disappears and a direct first-order transition between s-WC and p-WC occurs, which approaches $\phi=0.5$ when $\tilde\lambda$ is large. This can be qualitatively understood by considering a single electron on site $j$, which feels a static mean-field potential approximated by $U(\br-\mathbf{R_j})=\kappa (\br-\mathbf{R_j})^2/2$ due to electrons on the other sites. The projected single-particle Schrödinger equation is hence (taking $R_j=0$)
\begin{equation}
 \left\{\frac{k^2}{2m}+\frac{\kappa}{2}
 \left[(i\bm\nabla_{\bk}+\bm{\mcA})^2+\operatorname{tr}g\right]
 \right\}w(\bk)=Ew(\bk),
 \label{eq:trap}
\end{equation}
which describes a 2-D momentum-space harmonic oscillator problem in the presence of vector potential $\bm{\mcA}$ and scalar potential $\operatorname{tr}g$. We note that this approach is only qualitative and it fails in the AHC region due to the topological obstruction to constructing exponentially localized Wannier orbitals \cite{Thouless1984,Brouder2007}. 

Due to the rotational symmetry, the solution takes form $w_l(\bk)=f_l(k)e^{il\alpha_\bk}$, where $\alpha_\bk$ is the polar angle and $f_l(k)$ can be taken real. In the limit $\lambda \rightarrow 0$ the QG contributions are perturbative, ordinary rotationally symmetric oscillator therefore selects $l=0$, so the trivial crystal is always an $s$-WC. In the strong-limit $\lambda \rightarrow \infty$, Eq. (\ref{eq:trap}) becomes an Aharanov-Bohm problem and with a repulsive $k$-space potential $\text{tr}  g\rightarrow (1-\phi^2)/k^2$. By standard separation of variables, one obtains the leading-order energy is
\begin{equation}
 E_{n,l}^\text{strong}=\omega_{\kappa} (2n+1+\sqrt{(l-\phi)^2+1-\phi^2}),
\end{equation}
where $\omega_\kappa=\sqrt{\kappa/m}$. Note that for $0\leq |\phi|<1/2$ this is minimized by $l=0$, whereas for $1/2<|\phi|\leq1$ it is minimized by $l=\pm 1$, which explains the existence of p-WC and its transition to s-WC at $|\phi|=1/2$ for large $\tilde\lambda$. This competition highlights the importance of independently tuning QM and Berry flux: as we show below, strong QM can drive hollowing within the $s$-WC, whereas a sufficiently large Berry flux may favor competing crystalline phases, preempting the hollow $s$-WC as the ground state.

\begin{figure}[t]
 \centering
 \includegraphics[width=0.48\textwidth]{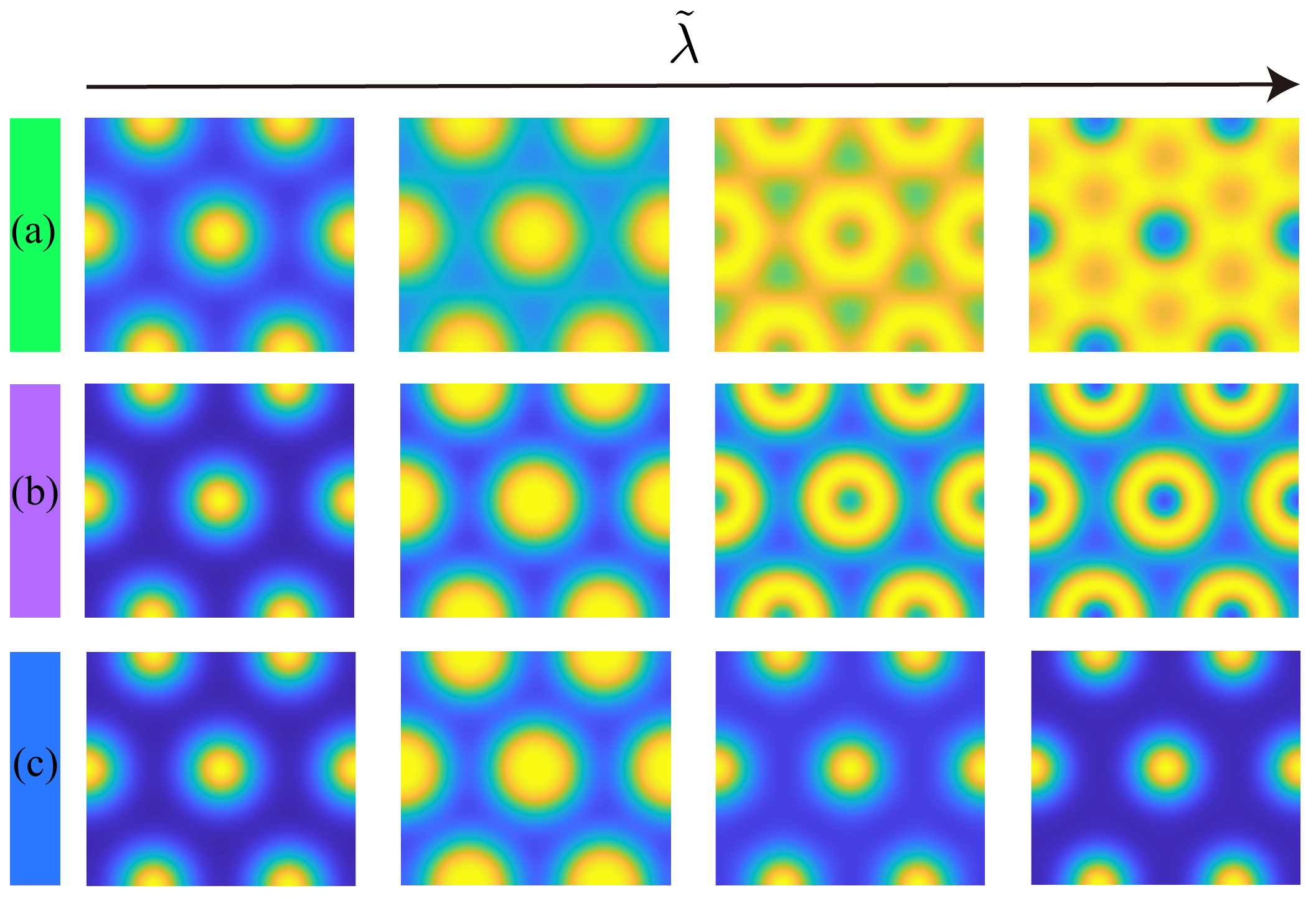}
 \caption{The density profile of the crystal at selected parameters (a) $r_s=9, \phi=0$ (b) $r_s=21, \phi=0$ (c) $r_s=21, \phi=1$, marked by colored hexagons in Fig. \ref{fig:phasediagram}. In each row, $\tilde\lambda$ increases from left to right. The color plot is determined by the density relative to the maximum density at each parameter, and yellow (blue) regions represent higher (lower) density.}
 \label{fig:density}
\end{figure}

\textit{Density profile and hollow crystal.---}
As demonstrated earlier, at $|\phi|<0.5$, the s-WC is always the ground state; however, it has a significantly different density profile in the weak and strong QG regimes. As shown in Fig. \ref{fig:density}(a)(b), the density distribution is concentrated near each site in the weak QG limit, but becomes significantly suppressed and develops a local minimum in the strong QG regime. Interestingly, at small $r_s=9$, the density rings of neighboring sites merge, resulting in a density profile with minima at the lattice sites and higher electron density in the interstitial regions. We refer to this s-WC with the new morphology as the hollow Wigner crystal (HWC), and emphasize that it is not a distinct thermodynamic phase as it adiabatically evolves to the weak QG s-WC without a phase transition. Such a ring structure has been well known to exist in higher Landau levels, which is attributed to the external magnetic field and oscillatory cyclotron form factors \cite{Koulakov1996,Fogler1996}. As we demonstrate below, here the hollowing is due to the strong QG and can persist even without a magnetic field or BC. 

The unique density profile of the HWC can be qualitatively understood with the local wave function $w(\bk)$ described by Eq. (\ref{eq:trap}), which determines the projected density in $k$-space
\begin{equation}
 \rho_l(\bq)=\int\frac{d^2k}{(2\pi)^2}
 w_l^*(\bk)w_l(\bk+\bq)F(\bk,\bk+\bq).
 \label{eq:packetdensity}
\end{equation}
When a rotationally symmetric density profile starts to develop a central minimum, the second derivative of real-space density at the origin
\begin{equation}
 \rho_l''(0)=-\frac{M_3}{4\pi},\qquad
 M_3\equiv\int_0^\infty dq\,q^3\rho_l(q),
 \label{eq:curvature}
\end{equation}
determines when the center starts to be depleted, and a central minimum requires a negative third moment $M_3<0$. 

\begin{figure}
    \centering 
    \subfigure[]{\includegraphics[height=0.25\textwidth]{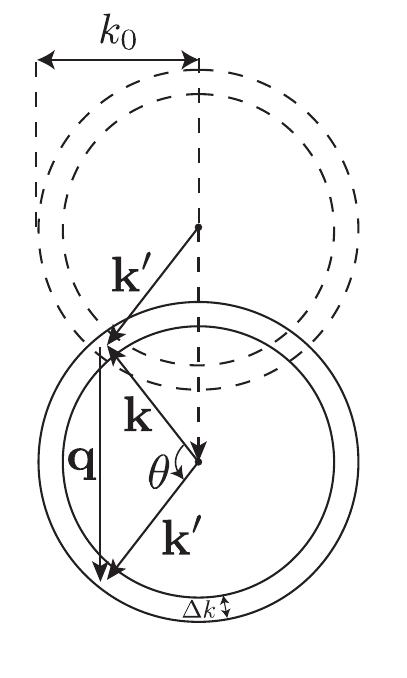} \label{fig:tworings}}
    \subfigure[]{\includegraphics[height=0.25\textwidth]{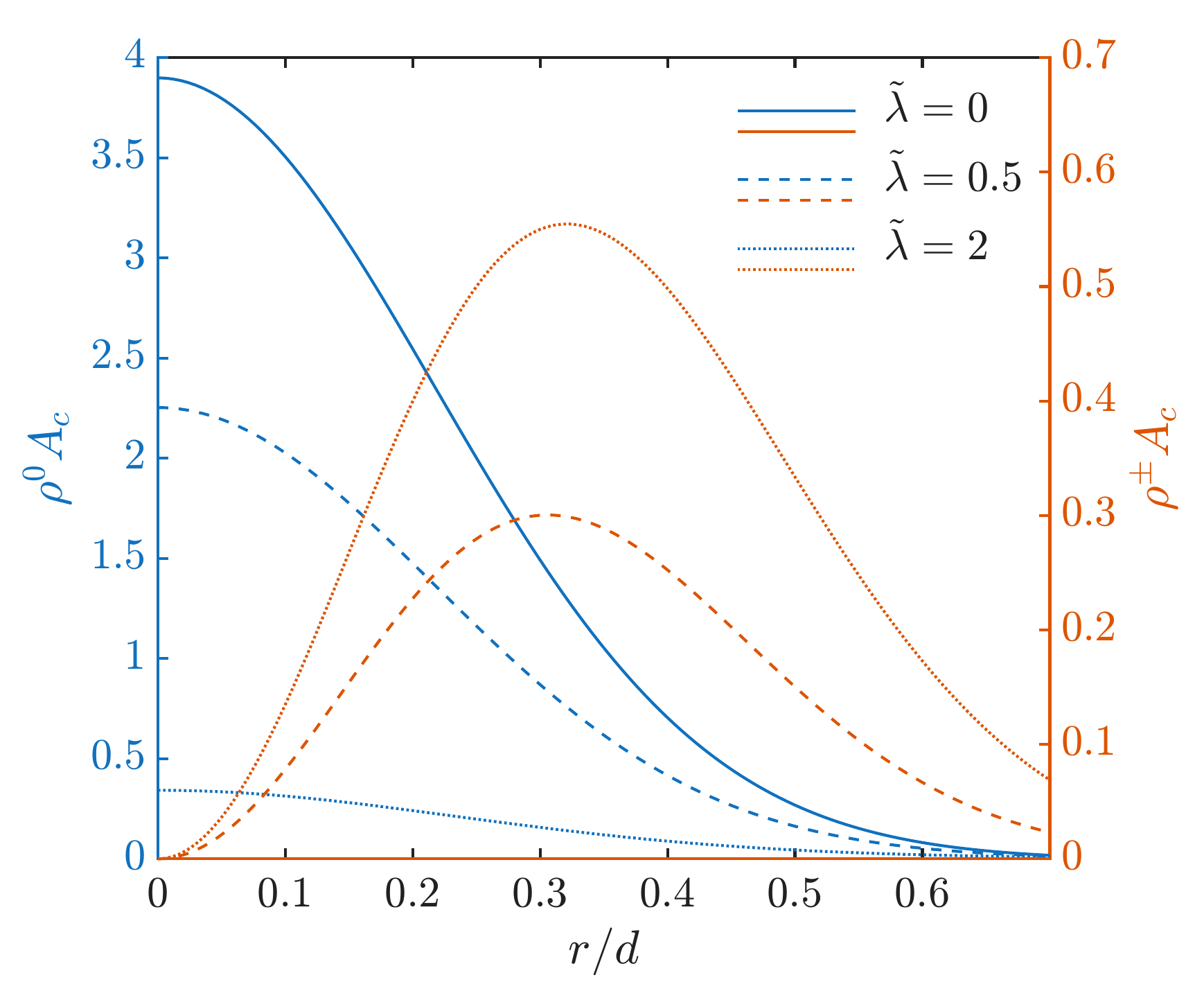}\label{fig:rhovsr}}
    \caption{(a) Schematics of the thin ring approximation. The dashed ring is obtained by translating the solid ring by $-\mathbf{q}$, and their overlap contributes to the integrand in Eq. (\ref{eq:packetdensity}). (b)The normalized density of the Wannier function $\rho^c A_c$ as a function of radius $r/d$, where $d$ is the triangular lattice constant and $A_c$ is the unit cell area. We take $r_s=9$, $\phi=0$, and plot the cut in the $x$-direction. }
\end{figure}

To simplify the analysis, we assume $f_l(k)=\sqrt{2\pi /(k_0\Delta k)}$ on a narrow ring of radius $k_0$ with width $\Delta k\ll k_0$ and zero elsewhere. The projected density $\rho_l(q)$ is thus proportional to the form factor $F(\bk,\bk')$ on the ring, where $|\bk|=|\bk'|=k_0$, $\theta$ is the angle between $\bk$ and $\bk'$ and $q=2k_0\sin\frac\theta2$, as depicted in Fig. \ref{fig:tworings}. Taking into account the shape of the overlap area, one can show that
\begin{equation}
 \rho_l(q)\simeq\frac{\Delta k\mathcal R_l(\theta)}{\pi k_0\sin\theta},
 \label{eq:ringdensity}
\end{equation}
where we have defined $\mathcal R_l(\theta)\equiv\operatorname{Re}[F(\bk,\bk')e^{il\theta}]$. Decomposing the form factor $F(\bk,\bk')=\sum_m p_m e^{im\theta}$, one obtains
\begin{align}
 M_3&\simeq\frac{2k_0^3\Delta k}{\pi}
 \int_0^\pi d\theta\,(1-\cos\theta)\mathcal R_l(\theta)\nonumber\\
 &=k_0^3\Delta k\left(2p_{-l}-p_{1-l}-p_{-1-l}\right).
 \label{eq:moment}
\end{align}
For the HWC we consider in this article, $l=0$ and the criterion becomes $p_1+p_{-1}>2p_0$, which further simplifies to
\begin{equation}
 \lambda^2k_0^2>2,
 \label{eq:modelthreshold}
\end{equation}
using $ p_0=1/(1+\eta)$ and $p_{\pm1}=\eta(1\mp\phi)/2(1+\eta)$ with $\eta=\lambda^2k_0^2$.

While the above analysis with thin-ring approximation seems to be oversimplifying and $k_0$ is set manually, it explains the origin of hollow structure, with $k_0$ being understood as the characteristic momentum scale of the wave function $w_l(\bk)$, typically the spread in $k$-space. Particularly, Eqs. (\ref{eq:ringdensity}) and (\ref{eq:moment}) indicate that the appearance of HWC is closely related to the angular structure of the form factor. For a general radial wave function $f_l(k)$, the hollowing criterion similar to Eq. (\ref{eq:moment}) can be derived without the thin-ring approximation, justifying the above treatment. A similar criterion can also be established for general rotationally symmetric QG models, although finite Berry flux may favor other crystalline states over the $s$-wave HWC (see End Matter) \footnote{We comment that in general HWC can have nonzero $l$; for example in Ref. \cite{Desrochers2026}, one of the ``Halo WC'' found in the $\lambda_N$-Jellium model also exhibits a hollow density profile but carries nonzero orbital angular momentum.}.

Due to the nontrivial band geometry in Eq.~(\ref{eq:spinor}), it is illuminating to resolve the three internal components $c=(0,+,-)$ of the localized wave function. We wannierize the occupied $C=0$ HF band using the Bloch wave function $\psi_k$
\begin{equation}
    |W_j\rangle=\frac{1}{\sqrt{N}}\sum_{\bk\in \text{BZ}} e^{-i(kR_j+\beta_\bk)} |\psi_\bk\rangle=\sum_\bk W_j(\bk)|\bk\rangle, \label{eq:wannierization}
\end{equation}
where $N$ is the number of sites in the supercell and a smooth Wannier gauge $\beta_\bk$ is chosen to minimize the orbital spread employing the standard approach \cite{Marzari1997}. The three-component real-space Wannier function $W_j(r)=(W_j^0(r),W_j^+(r),W_j^-(r))^T$ can then be found using
\begin{equation}
W_j(\br)=\int\frac{d^2k}{(2\pi)^2}
W_j(\bk)e^{i\bk\cdot\br}u_\bk^0.
\label{eq:wannier_real}
\end{equation}
The component-resolved density of the Wannier orbital can thus be calculated using $\rho^c=|W_0^{c}|^2$, which is plotted in Fig. \ref{fig:rhovsr}. At $\tilde \lambda=0$, the density is solely contributed by $\rho^0$, which has the ordinary center-peaked shape. At nonzero $\tilde \lambda$, $\rho^{\pm}$, which is peaked at finite radius, starts to contribute and finally dominates in the strong QG regime, explaining the hollow structure seen in Fig. \ref{fig:density}.

The component-resolved density can be understood analytically. Assuming $W_0(\bk)=w_l(\bk)=f_l(k)e^{il\alpha_\bk}$, with Eqs. (\ref{eq:wannier_real}) and (\ref{eq:spinor}), one obtains
\begin{equation}
w_l(\br)=
\begin{pmatrix}
e^{il\alpha_\br}\chi_l(r)\\
ia e^{i(l+1)\alpha_\br}
\left(\chi_l'-\frac{l}{r}\chi_l\right)\\
ib e^{i(l-1)\alpha_\br}
\left(\chi_l'+\frac{l}{r}\chi_l\right)
\end{pmatrix},
\label{eq:omegar}
\end{equation}
where $\chi_l'=\partial_r\chi_l$, $\alpha_\br$ is the real-space polar angle and
\begin{equation}
\chi_l(r)=i^l\int_0^\infty\frac{k\,dk}{2\pi}
\frac{f_l(k)}{\sqrt{1+\lambda^2 k^2}}
J_l(kr),
\end{equation}
$J_l(x)$ is the Bessel function. The component-resolved density is therefore
\begin{equation}
\rho_l^c(r)=
\left\{
\begin{array}{ll}
|\chi_l|^2, & c=0,\\
a^2|\chi_l'-l\chi_l/r|^2, & c=+,\\
b^2|\chi_l'+l\chi_l/r|^2, & c=-.
\end{array}
\right.
\label{eq:realspace_density}
\end{equation}
As $J_l(x)\sim x^l$, while the exact shape of $\rho_0^{\pm}$ depends on $f_l(k)$, they must vanish at the origin and be peaked at finite $r$. Therefore, the $c=0$ component dominates in the weak QG regime, responsible for the ordinary center-peaked density; in the strong QG regime, $\rho_0^{\pm}$ becomes more important, resulting in a ring-shaped density. The characteristic of the channel-resolved density (\ref{eq:realspace_density}) also explains the hollowing seen before.

\textit{Anomalous direct exchange.---}
As seen before, the hollow crystal is built from spatially broad
single-particle orbitals, which may enhance the exchange interaction between neighboring sites. 
Motivated by this, we calculate the intersite exchange energy (per electron)  using the Wannier orbitals found previously\footnote{We find that the Wannier gauge $\beta_\bk$ we choose in Eq. (\ref{eq:wannierization}) not only minimizes orbital spread but also $\varepsilon_{exc}$. The reported $\varepsilon_{exc}$ therefore serves as a conservative estimate of the lower bound of the exchange coupling experienced by a localized electron.}
\begin{equation}
    \varepsilon_{exc}=-e_F-\frac{U_W}{2}=\frac{1}{2}\sum_{j\neq 0} K_{0j}
\end{equation}
where $e_F<0$ is the total Fock energy per electron, $U_W=K_{00}>0$ is the self-energy, and the exchange matrix element reads
\begin{equation}
 K_{ij}=\int d^2r\,d^2r'\,
 [W_i^\dagger(\br)W_j(\br)]V(\br-\br')
 [W_j^\dagger(\br')W_i(\br')].
 \label{eq:Kij}
\end{equation}
For spinful electrons, $\Delta E_\text{flip}=2\varepsilon_{exc}$ thus serves as an estimate for the energy change when flipping a single lattice spin in a fully polarized Hollow s-WC. 

\begin{figure}[!t]
 \includegraphics[width=0.45 \textwidth]{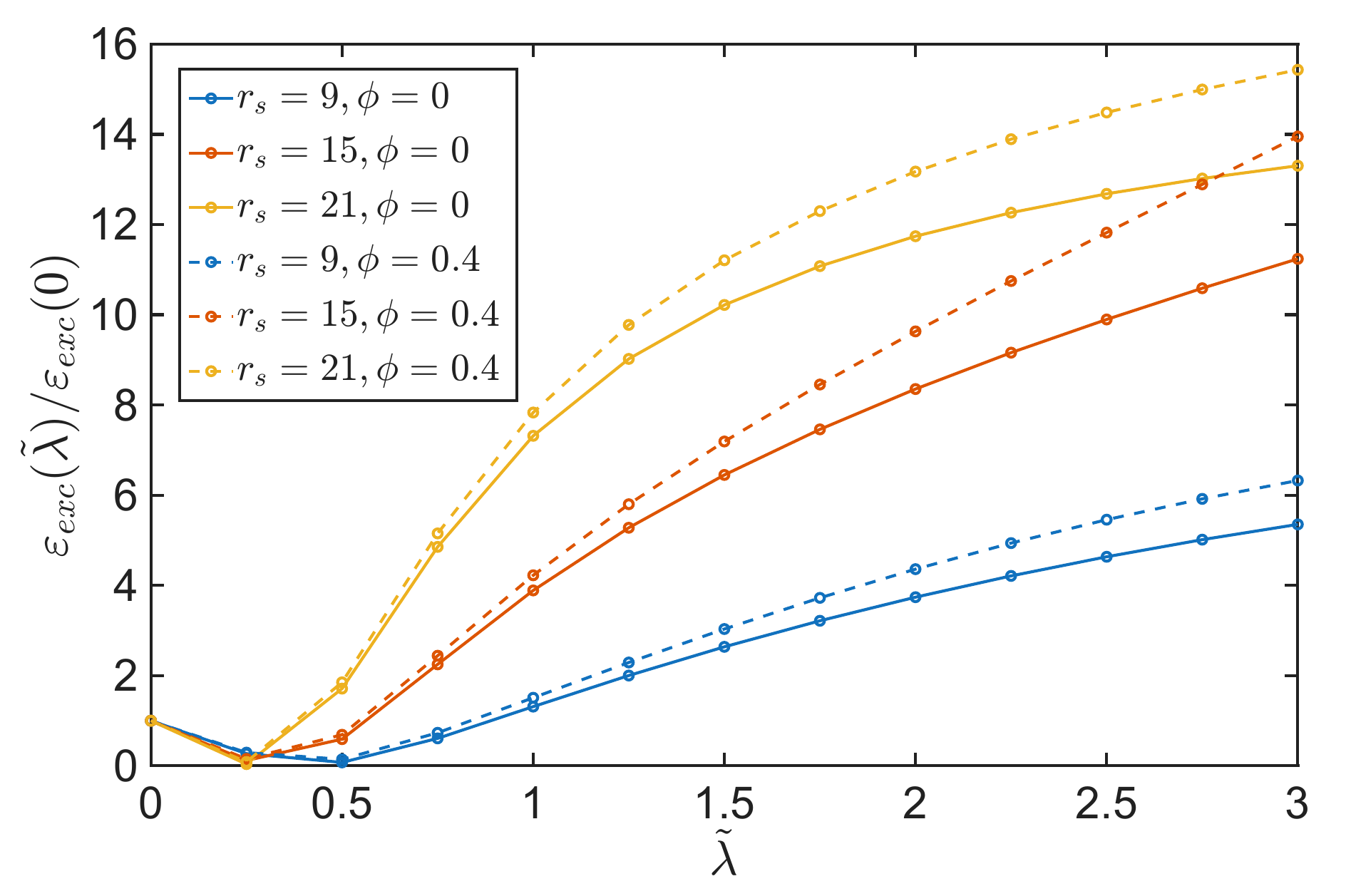}
\caption{The intersite exchange energy ratio $\varepsilon_{exc}(\tilde \lambda)/\varepsilon_{exc}(0)$ for different $r_s$ and $\phi$. The exchange energy first develops a dip around $0.25-0.5$, then is strongly enhanced in the strong QG regime.}
 \label{fig:exchange}
\end{figure}
As seen from Fig.~\ref{fig:exchange}, the exchange energy $\varepsilon_{exc}$ is indeed significantly enhanced for the s-WC in the strong QG regime, ranging from several times to an order of magnitude. This is attributed to the large real-space overlap between adjacent orbitals in the strong QG regime. Interestingly, we also find that $\varepsilon_{exc}$ becomes strongly suppressed at intermediate $\tilde \lambda\sim 0.25-0.5$, which is due to the strong destructive phase interference between different exchange channels, as demonstrated below. To make this mechanism explicit, we define the overlap function $\xi_{ij}(\br)=W_i^\dagger(\br)W_j(\br)$ and use its value at the midpoint $\br_{m}=(\mathbf{R_i}+\mathbf{R_j})/2$ to estimate the exchange matrix element (\ref{eq:Kij}) (see End Matter for justification), giving
\begin{equation}
    K_{ij}\sim \Lambda^3|\xi_{ij}(\br_{m})|^2,
\end{equation}
where $\Lambda$ is the length scale of the non-vanishing region of $|\xi_{ij}(\br)|$. As seen from Eq. (\ref{eq:omegar}), $W_i^0(\br_{m})=W_j^0(\br_{m})$ and $W_i^\pm(\br_{m})=-W_j^\pm(\br_{m})$, we obtain
\begin{equation}
    \xi_{ij}(\br_{m})=|W_i^0(\br_{m})|^2-|W_i^+(\br_{m})|^2-|W_i^-(\br_{m})|^2.
\end{equation}
As demonstrated before, the first term dominates in the weak QG regime while in the strong QG regime the last two terms win. Hence at the crossover between these two regimes $\xi_{ij}(\br_{m})$ must vanish, explaining the dip in the exchange energy $\varepsilon_{exc}$.

\textit{Discussion.---}
In summary, we have introduced a continuum model with separately tunable QM and Berry flux and investigated its Hartree-Fock phase diagram. The interplay between strong QM and the nontrivial angular structure of the form factor redistributes wave-function weight among different channels, leading to the formation of HWC. In our minimal model, this orbital reconstruction also produces a pronounced suppression of direct Coulomb exchange through destructive interference between different channels in the orbital overlap. The strong delocalization of orbitals further leads to a significant enhancement of direct exchange in the strong-metric regime.

Our findings suggest several interesting directions for future investigations. Enhanced direct exchange may lead to much stronger magnetism in HWCs compared to conventional WCs \cite{Care1975,Herman1984,Zhu1995,Kim2026,Esterlis2025}, although the net magnetic coupling also depends on superexchange \cite{Flambaum1999} and multiparticle ring exchange \cite{Roger1984,Bernu2001}. Near the crossover, the sign reversal of intersite orbital overlap suggests that gauge-invariant products around closed loops, such as $\xi_{ij}\xi_{jk}\xi_{ki}$, may also change sign, potentially affecting higher-order cyclic exchange interactions. The characteristic hollow density profile could be directly visualized by scanning tunneling microscopy through spatially resolved measurements of the occupied-state spectral weight. Hollow density profiles may also favor unconventional interlayer arrangements, for example between conventional and hollow crystals, by reducing Coulomb repulsion \cite{Goldoni1996,Zhuang2025}. More broadly, our three-band model provides a minimal platform for exploring diverse QM effects in other correlated states, advancing our understanding of the interplay between correlation and QG.

\textit{Acknowledgements.---}
We thank Ilya Esterlis for useful discussion and feedback. This work was supported by the National Natural Science Foundation of China (Grant No. 12604224) and the Scientific Foundation for Youth Scholars of Shenzhen University (Grant No. 868-000001033517).

\nocite{Dong2026,Han2026}
\bibliography{reference}


\clearpage
\onecolumngrid
\begin{center}
    {\large\bfseries End Matter}
\end{center}
\vspace{4pt}
\twocolumngrid

\appendix*
\setcounter{equation}{0}
\section{Hartree--Fock theory}

We solve the projected Coulomb Hamiltonian using self-consistent Hartree--Fock (HF) theory. For a crystal with reciprocal vectors $\mathbf G$, the density matrix satisfies
$\langle c^\dagger_{\mathbf k}c_{\mathbf k+\mathbf q}\rangle
=n_{\mathbf k}(\mathbf G)\delta_{\mathbf q,\mathbf G}$,
and the projected density is
\begin{equation}
\rho_{\mathbf G}
=\sum_{\mathbf k}F(\mathbf k,\mathbf k+\mathbf G)
n_{\mathbf k}(\mathbf G).
\label{em:rhoG}
\end{equation}
The Hartree and Fock contributions to the HF Hamiltonian read
\begin{align}
H_{\rm H}
={}&\frac{1}{A}\sum_{\mathbf G\ne0,\mathbf k}
V(G)F(\mathbf k,\mathbf k-\mathbf G)
\rho_{\mathbf G}c^\dagger_{\mathbf k}
c_{\mathbf k-\mathbf G},
\nonumber\\
H_{\rm F}
={}&-\frac{1}{A}\sum_{\mathbf G,\mathbf k,\mathbf k'}
V(\mathbf k'-\mathbf k-\mathbf G)
\nonumber\\
&\times F(\mathbf k,\mathbf k'-\mathbf G)
F(\mathbf k',\mathbf k+\mathbf G)
n_{\mathbf k}(\mathbf G)
c^\dagger_{\mathbf k'}c_{\mathbf k'-\mathbf G}.
\label{em:HFfields}
\end{align}
The uniform positive background removes the $\mathbf G=0$ Hartree term. 
The total HF energy, including the double-counting correction, is
\begin{equation}
E_{\rm HF}
=\sum_{\mathbf k}\epsilon_{\mathbf k}n_{\mathbf k}(0)
+\frac12\langle H_{\rm H}+H_{\rm F}\rangle.
\label{em:HFenergy}
\end{equation}
At each reduced crystal momentum, we diagonalize the HF Hamiltonian in the plane-wave basis $\{\mathbf k+\mathbf G\}$ and update the density matrix from the occupied HF Bloch states until self-consistency. The physical density is
$n(\mathbf r)=A^{-1}\sum_{\mathbf G}\rho_{\mathbf G}e^{i\mathbf G\cdot\mathbf r}$.

We compare triangular and square lattices with one electron per primitive cell, using up to 29 reciprocal lattice vectors and $24\times24$ momentum meshes. The optimal damping algorithm (ODA) and direct inversion in the iterative subspace (DIIS) accelerate convergence. We subtract the Madelung correction and compare solutions from multiple initial density matrices with different lattice and angular-momentum characters.

\section{Phase characterization}

The crystalline phases can be characterized by their rotation eigenvalues at high-symmetry momenta \cite{Fang2012}. For a rotationally symmetric crystal, the occupied Hartree--Fock (HF) Bloch state at a rotation-invariant momentum $\mathbf k$ satisfies
\begin{equation}
\hat C_n|\psi_{\mathbf k}\rangle
=\eta_n(\mathbf k)|\psi_{\mathbf k}\rangle,
\end{equation}
where $\eta_n(\mathbf k)$ is an $n$th root of unity. For the triangular lattice, the relevant rotation eigenvalues are $\eta_6(\Gamma)$, $\eta_3(K)$, and $\eta_2(M)$, while for the square lattice they are $\eta_4(\Gamma)$, $\eta_4(M)$, and $\eta_2(X)$.

For the topologically trivial triangular WCs considered in the main text, the $s$-WC has trivial rotation eigenvalues,
\begin{equation}
\eta_6(\Gamma)=\eta_3(K)=\eta_2(M)=1.
\end{equation}
In contrast, the $p$-WC carries one unit of orbital angular momentum. For $l=\pm1$, its rotation eigenvalues are
\begin{equation}
\eta_6(\Gamma)=e^{\mp i\pi/3},\qquad
\eta_3(K)=e^{\mp i2\pi/3},\qquad
\eta_2(M)=-1.
\end{equation}

The rotation eigenvalues also constrain the Chern number $C$ of the occupied HF band. For a spinless system with $C_6$ or $C_4$ symmetry, respectively,
\begin{align}
e^{i2\pi C/6}
&=\eta_6(\Gamma)\eta_3(K)\eta_2(M),\\
e^{i2\pi C/4}
&=\eta_4(\Gamma)\eta_4(M)\eta_2(X).
\end{align}
Both the $s$-WC and $p$-WC are topologically trivial ($C=0$), whereas the anomalous Hall crystals (AHCs) identified in this work have $|C|=1$. For $\phi>0$, the AHC appearing in the phase diagram has $\eta_6(\Gamma)=e^{-i\pi/3}$ and $\eta_3(K)=\eta_2(M)=1$, consistent with its Chern number $C=-1$.

\begin{figure}[t]
    \centering
    \includegraphics[width=0.49\textwidth]{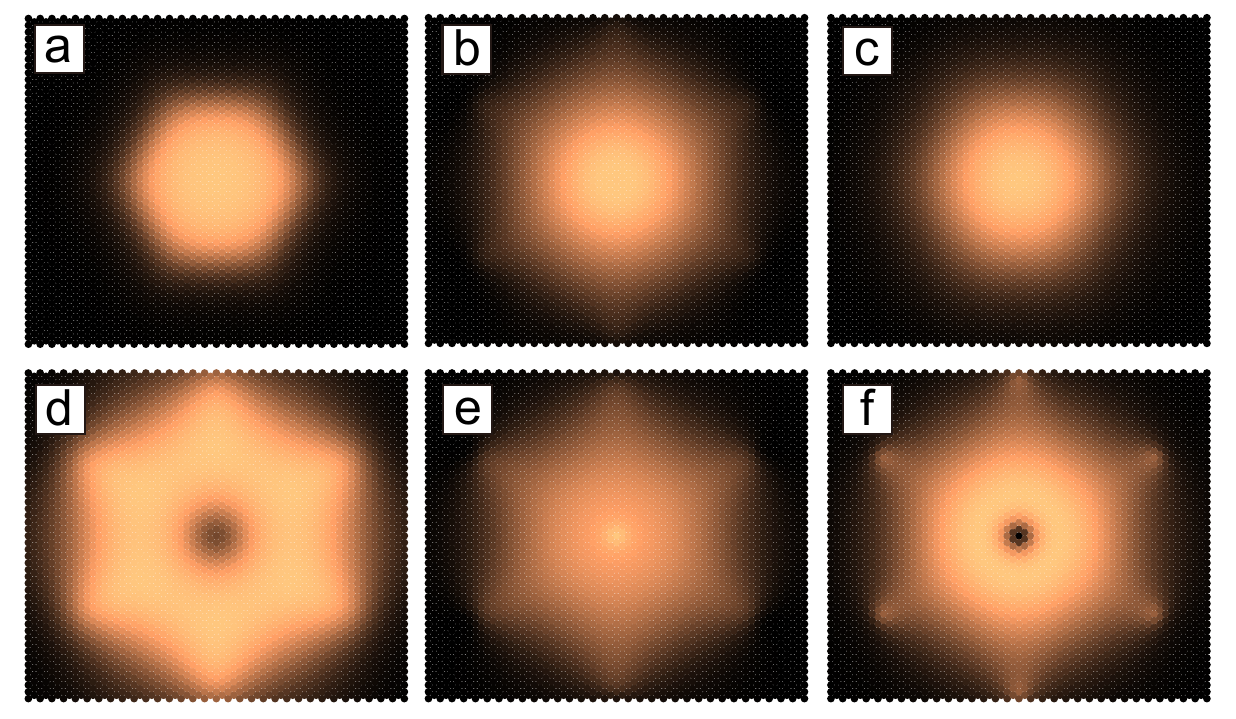}
    \caption{Momentum-space occupation $n(\mathbf k)=\langle c^\dagger_{\mathbf k}c_{\mathbf k}\rangle$ at $(r_s,\phi,\tilde\lambda)=$ (a) $(9,0,0)$, (b) $(9,0,2)$, (c) $(21,0,0)$, (d) $(21,0,2)$, (e) $(21,1,1)$, and (f) $(21,1,2)$. Panels (a) to (d) correspond to $s$-WCs, while (e) and (f) correspond to an AHC and a $p$-WC, respectively.}
    \label{fig:kspaceocc}
\end{figure}

The distinction between $s$-WC and $p$-WC is also evident in their momentum-space occupations, $n(\mathbf k)=\langle c^\dagger_{\mathbf k}c_{\mathbf k}\rangle$, shown in Fig.~\ref{fig:kspaceocc}. For the $s$-WC, the occupation at $\mathbf k=0$ is generally nonzero, even when its real-space density becomes hollow. In contrast, the nontrivial rotation character of the $p$-WC requires $n(\mathbf k=0)=0$, consistent with Fig.~\ref{fig:kspaceocc}(f).

\section{Hollowing criterion for rotationally symmetric bands}

Consider an $N$-component Hamiltonian with continuous rotational symmetry, satisfying
\begin{equation}
H(R_\alpha\mathbf k)
=U_\alpha H(\mathbf k)U_\alpha^\dagger,
\quad
U_\alpha=\sum_a e^{im_a\alpha}|a\rangle\langle a|,
\end{equation}
where $R_\alpha$ denotes a counterclockwise rotation in momentum space, and $U_\alpha$ acts on the internal orbital space. For an isolated, nondegenerate band, a rotation-adapted eigenvector can be expressed as
$|u_{k,\theta}\rangle=\sum_a c_a(k)e^{im_a\theta}|a\rangle$. Its form factor can be expanded as
$F(k,k',\vartheta)=\sum_m F_m(k,k')e^{im\vartheta}$,
where $F_m(k,k')=\sum_{a:m_a=m}c_a^*(k)c_a(k')$
and $\vartheta=\theta_{\mathbf k'}-\theta_{\mathbf k}$. For the scalar envelope $w_l(\mathbf k)=f(k)e^{il\theta}$, integrating over $\vartheta$ in the moment $M_3$ defined in the main text gives the exact relation
\begin{equation}
M_3\big|_l
=2\pi\left(2P_{-l}-Q_{1-l}-Q_{-1-l}\right),
\label{em:generalM3}
\end{equation}
where $d\mu(k)=k\,dk/(2\pi)$ and
\begin{align}
P_m
&=\int d\mu(k)d\mu(k')f^*(k)f(k')\frac{k^2+k'^2}{2}F_m(k,k'),
\nonumber\\
Q_m
&=\int d\mu(k)d\mu(k')f^*(k)f(k')
kk'F_m(k,k').
\label{em:PQ}
\end{align}
Thus, only three neighboring angular harmonics of the form factor contribute to the quadratic curvature of the real-space density, regardless of radial width.
\begin{table}[hb]
\caption{Angular weights and thin-ring $s$-wave hollowing criteria for representative models. Here $\eta=\lambda^2k_0^2$, $x_0=\mathcal Bk_0^2/2$, and $|c_0|>|c_1|$ for the positive-mass Dirac conduction band ($N=1$).}
\label{tab:othermodels}
\centering
\begingroup
\setlength{\tabcolsep}{2pt}
\newcommand{\EMcell}[3]{%
    \parbox[c][#1][c]{#2}{\centering #3}}

\begin{ruledtabular}
\begin{tabular}{ccc}

\EMcell{8ex}{0.29\columnwidth}{Model}
&
\EMcell{8ex}{0.43\columnwidth}{Angular weights $p_m$}
&
\EMcell{8ex}{0.21\columnwidth}{Hollowing criterion}
\\
\hline

\EMcell{11ex}{0.29\columnwidth}{
$\lambda$-Jellium \cite{Soejima2025}}
&
\EMcell{11ex}{0.43\columnwidth}{
$\displaystyle
\begin{gathered}
p_0=\frac{1}{1+\eta}\\
p_{-1}=\frac{\eta}{1+\eta}
\end{gathered}$}
&
\EMcell{11ex}{0.21\columnwidth}{
$\eta>2$}
\\
\hline

\EMcell{13ex}{0.29\columnwidth}{
Chiral two-band\\
($N=1$: Dirac;\\
$N\geq2$: RMG) \cite{Zeng2024}}
&
\EMcell{13ex}{0.43\columnwidth}{
$\displaystyle
\begin{gathered}
p_0=|c_0(k_0)|^2\\
p_{-N}=|c_N(k_0)|^2
\end{gathered}$}
&
\EMcell{13ex}{0.21\columnwidth}{
Not satisfied}
\\
\hline

\EMcell{11ex}{0.29\columnwidth}{
Uniform-BC ideal band\\
\cite{Tan2024,Desrochers2026}}
&
\EMcell{11ex}{0.43\columnwidth}{
$\displaystyle
p_{-n}=e^{-x_0}\frac{x_0^n}{n!}$}
&
\EMcell{11ex}{0.21\columnwidth}{
$\mathcal Bk_0^2>4$}
\\

\end{tabular}
\end{ruledtabular}
\endgroup
\end{table}

For our three-band model,
$F_0=(DD')^{-1}$ and
$F_{\pm1}=\lambda^2(1\mp\phi)kk'/(2DD')$,
where $D(k)=\sqrt{1+\lambda^2k^2}$ and
$D'=D(k')$.
For a real, nonnegative radial envelope $f(k)$, define
$A_j=\int d\mu(k)k^jf(k)/D(k)$.
Substitution into Eq.~\eqref{em:PQ} gives
\begin{align}
P_0&=A_0A_2,
&
P_{\pm1}&=\frac{\lambda^2(1\mp\phi)}{2}A_1A_3,
\nonumber\\
Q_0&=A_1^2,
&
Q_{\pm1}&=\frac{\lambda^2(1\mp\phi)}{2}A_2^2.
\label{em:PQthree}
\end{align}
According to the definitions in the main text
\begin{equation}
p_0(k)=\frac{1}{1+\lambda^2k^2},
\quad
p_{\pm1}(k)=
\frac{\lambda^2k^2(1\mp\phi)}
{2(1+\lambda^2k^2)}.
\end{equation}
Consequently,
\begin{align}
\frac{Q_{\pm1}}{P_0}
&=\frac{p_{\pm1}(k_s)}{p_0(k_s)},
&
k_s^2&=\frac{A_2}{A_0},
\nonumber\\
\frac{P_{\mp1}}{Q_0}
&=\frac{p_{\mp1}(k_p)}{p_0(k_p)},
&
k_p^2&=\frac{A_3}{A_1}.
\end{align}
For $l=0,\pm1$, the exact moment therefore reduces to
\begin{equation}
M_3\big|_l=C_l
\left[
2p_{-l}(k_0)-p_{1-l}(k_0)-p_{-1-l}(k_0)
\right],
\label{em:exactM3}
\end{equation}
with $k_0=k_s$ for $l=0$ and $k_0=k_p$ for $l=\pm1$, and the positive prefactor
\begin{equation}
C_l=2\pi(1+\lambda^2k_0^2)
\begin{cases}
A_0A_2,&l=0,\\
A_1^2,&l=\pm1.
\end{cases}
\end{equation}
Hence, we have proved the form of Eq. (\ref{eq:moment}) holds for general $f(k)$, with the characteristic momentum $k_0$ in the main text being identified with $k_s$ or $k_p$.

Table~\ref{tab:othermodels} summarizes the application of the hollowing criterion to other quantum-geometric models. Satisfying the criterion, however, does not guarantee that the hollow $s$-WC is energetically favored over competing crystalline states.

\section{Numerical examination of the midpoint approximation}

\begin{figure}[h]
\includegraphics[width=0.45\textwidth]{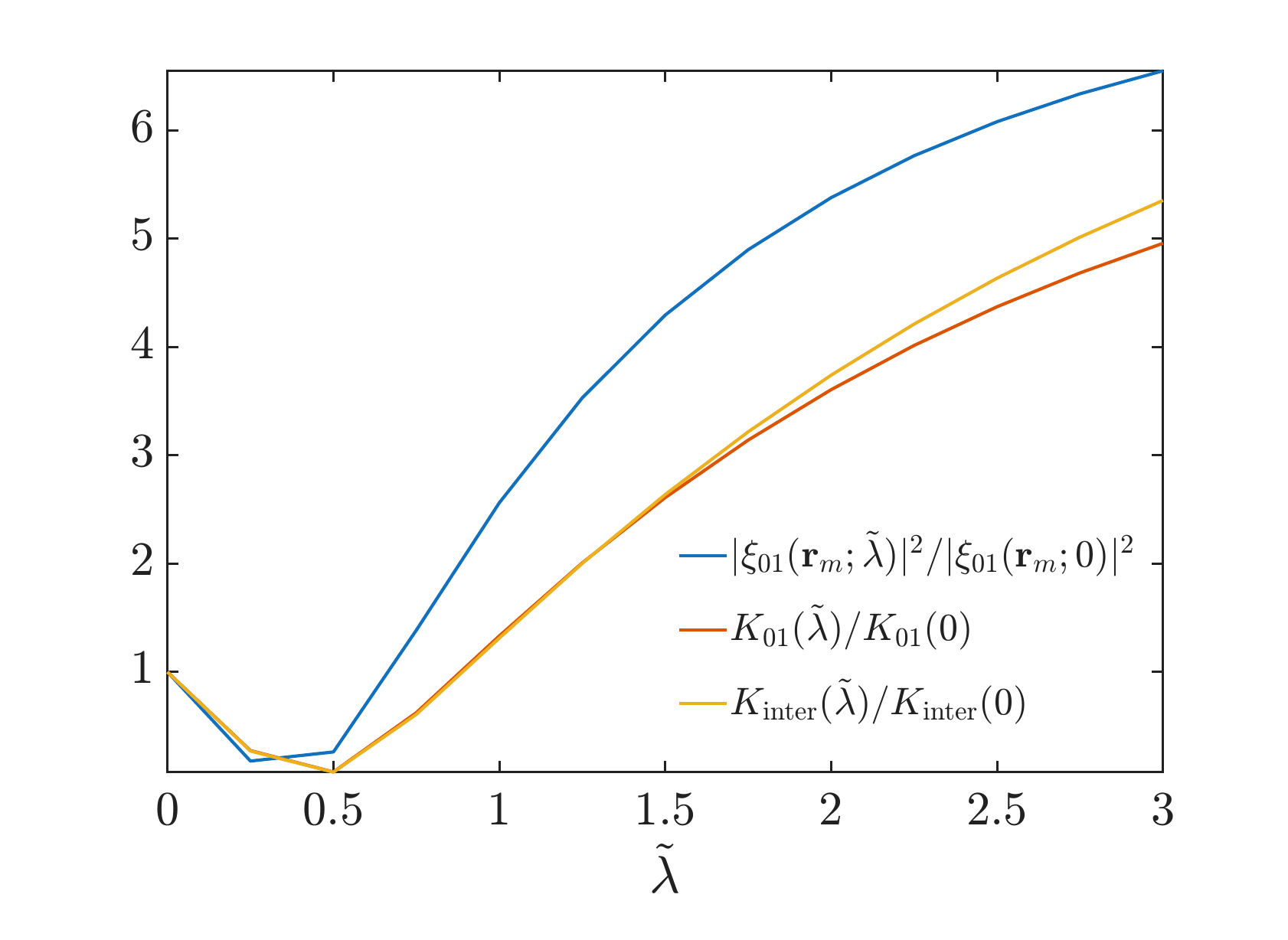}
\caption{Comparison between the midpoint overlap
$|\xi_{01}(\mathbf r_m)|^2$, exact nearest-neighbor
exchange $K_{01}$, and total intersite exchange
$K_{\rm inter}=\sum_{j\ne0}K_{0j}$ for $\phi=0$
at $r_s=9$. All quantities are normalized to their values
at $\tilde\lambda=0$.}
\label{fig:midpoint_validation}
\end{figure}

Figure~\ref{fig:midpoint_validation} compares $|\xi_{01}(\mathbf r_m)|^2$ with the exact $K_{01}$ evaluated from Eq.~(\ref{eq:Kij}). The midpoint approximation qualitatively reproduces the behavior of $K_{01}$, which dominates the total intersite exchange $K_{\rm inter}$.

\end{document}